\documentclass[english,showpacs,preprint,superscriptaddress]{revtex4-1}
\usepackage{lmodern}
\usepackage[T1]{fontenc}
\usepackage{babel}
\usepackage{mathtools}
\usepackage{amsmath}
\usepackage{amsthm}
\usepackage{amssymb}
\usepackage[unicode=true,
 bookmarks=false,
 breaklinks=false,pdfborder={0 0 1},backref=false,colorlinks=false]
 {hyperref}

\makeatletter

\usepackage{amsthm}\usepackage{latexsym}\usepackage{bm}\usepackage{amsfonts}

\allowdisplaybreaks
\usepackage{babel}
\usepackage{babel}

\usepackage{babel}

\makeatother

\begin{document}
\title{A gauge-symmetrization method for energy-momentum tensors in high-order
electromagnetic field theories}
\author{Peifeng Fan}
\affiliation{Key Laboratory of Optoelectronic Devices and Systems, College of Physics
and Optoelectronic Engineering, Shenzhen University, Shenzhen 518060,
China}
\affiliation{Advanced Energy Research Center, Shenzhen University, Shenzhen 518060,
China}
\author{Jianyuan Xiao}
\affiliation{School of nuclear science and technology, University of Science and
Technology of China, Hefei, Anhui 230026, China}
\author{Hong Qin}
\affiliation{Princeton Plasma Physics Laboratory, Princeton University, Princeton,
NJ 08543, USA}
\begin{abstract}
For electromagnetic field theories, canonical energy-momentum conservation
laws can be derived from the underpinning spacetime translation symmetry
according to the Noether procedure. However, the canonical Energy-Momentum
Tensors (EMTs) are neither symmetric nor gauge-symmetric (gauge invariant).
The Belinfante-Rosenfeld (BR) method is a well-known procedure to
symmetrize the EMTs, which also render them gauge symmetric for first-order
field theories. High-order electromagnetic field theories appear in
the study of gyrokinetic systems for magnetized plasmas and the Podolsky
system for the radiation reaction of classical charged particles.
For these high-order field theories, gauge-symmetric EMTs are not
necessarily symmetric and vice versa. In the present study, we develop
a new gauge-symmetrization method for EMTs in high-order electromagnetic
field theories. The Noether procedure is carried out using the Faraday
tensor $F_{\mu\nu}$, instead of the 4-potential $A_{\mu}$, to derive
a canonical EMT $T_{\text{N}}^{\mu\nu}$. We show that the gauge-dependent
part of $T_{\text{N}}^{\mu\nu}$ can be removed using the displacement-potential
tensor $\mathcal{F}^{\sigma\mu\nu}\equiv\mathcal{D}^{\sigma\mu}A^{\nu}/4\pi$,
where $\mathcal{D}^{\sigma\mu}$ is the anti-symmetric electric displacement
tensor. This method gauge-symmetrize the EMT without necessarily making
it symmetric, which is adequate for applications not involving general
relativity. For first-order electromagnetic field theories, such as
the standard Maxwell system, $\mathcal{F}^{\sigma\mu\nu}$ reduces
to the familiar BR super-potential $\mathcal{S}^{\sigma\mu\nu}$,
and the method developed can be used as a simpler procedure to calculate
$\mathcal{S}^{\sigma\mu\nu}$ without employing the angular momentum
tensor in 4D spacetime. When the electromagnetic system is coupled
to classical charged particles, the gauge-symmetrization method for
EMTs is shown to be effective as well. 
\end{abstract}
\maketitle

\section{introduction}

In classical field theories, one can derive canonical Energy-Momentum
Tensors (EMTs) $T_{\text{N}}^{\mu\nu}$ from the underpinning spacetime
translation symmetry using the Noether procedure \citep{Noether1918}.
However, for classical systems of electromagnetic field, the canonical
EMTs are neither symmetric with respect to tensor indices nor electromagnetic
gauge invariant. Gauge dependence is un-physical, and non-symmetric
EMT is not consistent with general relativity. In the present study,
we will call a EMT symmetric if it is symmetric with respect to tensor
indices, and gauge symmetric if it is gauge invariant. To date, much
effort has been focused on symmetrizing the EMTs (with respect to
tensor indices), while constructing gauge-symmetric EMTs is oftentimes
a challenging task for general systems \citep{Babak1999,Gratus2012,Arminjon2016,Inglis2016,Jimenez2018,Ilin2020}. 

The first method for symmetrizing EMTs was discovered by Belinfante
\citep{Belinfante1939,Belinfante1940} and Rosenfeld \citep{Rosenfeld1940},
who added a divergence-free tensor $\partial_{\sigma}\mathcal{S}^{\sigma\mu\nu}$
to obtain a symmetric EMT, i.e., 
\begin{align}
 & T_{\mathrm{BR}}^{\mu\nu}=T_{\text{N}}^{\mu\nu}+\partial_{\sigma}\mathcal{S}^{\sigma\mu\nu},\label{eq:T_BR}\\
 & \partial_{\mu}\partial_{\sigma}\mathcal{S}^{\sigma\mu\nu}=0.
\end{align}
Here, $T_{\mathrm{BR}}^{\mu\nu}$ is Belinfante-Rosenfeld (BR) EMT,
and $\mathcal{S}^{\sigma\mu\nu}$ is known as BR super-potential that
depends on the angular momentum tensor and is anti-symmetric with
respect to $\sigma$ and $\mu$ {[}see Eq.\,(\ref{eq:cov-BR-tensor}){]}.
General relativity suggests another method to generate symmetric EMTs
\citep{Hawking&Ellis1973,Landau1975}, which was modified by Gotay
and Marsden, who employed constraints to define symmetric EMTs \citep{Gotay1992,Lopez2007a}.
The relations between these three types of symmetric EMTs have been
discussed in the literature \citep{Zhang2005,Ilin2019,Baker2021}.

In many systems, including the standard Maxwell system (\ref{eq:Maxwell-Lagrangian}),
the symmetrization of $T_{\text{N}}^{\mu\nu}$ also renders it gauge
symmetric. But for general electromagnetic field theories with high-order
field derivations, symmetry with respect to tensor indices in general
does not imply gauge symmetry and vice versa. High-order electromagnetic
field theories appear in the study of gyrokinetic systems \citep{Qin2005,Qin2007a,Fan2020}
for magnetized plasmas and the Podolsky system \citep{Bopp1940,Podolsky1942}
for the radiation reaction of classical charged particles. In the
present study, we propose a new method to gauge-symmetrize the canonical
EMTs $T_{\text{N}}^{\mu\nu}$ in general electromagnetic field theories
with high-order field derivations. Our method removes the gauge dependence,
but does not necessarily symmetrize the EMTs. In applications that
don't involve general relativity, gauge-symmetrized EMTs are adequate. 

We first reformulate the equation of motion for the field by the variational
principle with respect to the Faraday tensor $F_{\mu\nu}$, instead
of the 4-potential $A^{\mu}$ as in the standard field theory. The
Euler-Lagrange (EL) equation is cast into an explicitly gauge-symmetric
form. The canonical EMT is then separated into a gauge-invariant part
and a gauge-dependent part, the later of which contains the anti-symmetric
electric displacement tensor $\mathcal{D}^{\mu\nu}$. We define a
super-potential $\mathcal{F}^{\sigma\mu\nu}\equiv\mathcal{D}^{\sigma\mu}A^{\nu}/4\pi$,
called displacement-potential tensor, whose divergence with respect
to the first index is divergence free with respect to the second index
and removes the gauge dependence in the canonical EMT. It is simpler
to calculate the displacement-potential tensor $\mathcal{F}^{\sigma\mu\nu}$
than the BR super-potential $\mathcal{S}^{\sigma\mu\nu}$, and the
former only gauge-symmetrizes the canonical EMT without render it
symmetric (with respect to tensor indices). For first-order electromagnetic
field theories, such as the standard Maxwell system, $\mathcal{F}^{\sigma\mu\nu}$
reduces to the familiar BR super-potential $\mathcal{S}^{\sigma\mu\nu}$,
and the method developed here can be used as a simpler procedure to
calculate $\mathcal{S}^{\sigma\mu\nu}$ without employing the angular
momentum tensor in 4D spacetime. In addition, when the electromagnetic
system is coupled with classical charged particles, we find that the
method is effective as well, even though the Lagrangian density is
not gauge symmetric in general.

The paper is organized as follows. In Sec.$\thinspace$\ref{sec:Gauge invariant conservation laws for electromagnetic system},
we describe the gauge-symmetrization method for the EMT in a general
high-order electromagnetic field theory, and highlight the difference
in comparison with the BR method using the example of the Podolsky
system \citep{Bopp1940,Podolsky1942}. Section \ref{sec:Particle-Field-system}
shows how the gauge-symmetrization method for the EMT works when the
electromagnetic system is coupled with classical charged particles.

\section{explicitly gauge-symmetric conservation laws for high-order electromagnetic
systems \label{sec:Gauge invariant conservation laws for electromagnetic system}}

\subsection{Explicitly gauge-symmetric Euler-Lagrange equation}

The Lagrangian density of a general electromagnetic system is written
as
\begin{equation}
\mathcal{L}_{F}=\mathcal{L}_{F}\left(x^{\mu},DA_{\mu},\cdots,D^{\left(n+1\right)}A_{\mu}\right),\label{eq:cov-EM-Lagrangian-A_=00005Cmu}
\end{equation}
where $A=\left(\varphi,-\boldsymbol{A}\right)$ is the 4-potential
and $D=\left(\left(1/c\right)\partial_{t},\boldsymbol{\nabla}\right)$
is the derivative operator over spacetime. The EL equation of the
Lagrangian density is 
\begin{equation}
E_{A}^{\mu}\left(\mathcal{L}_{F}\right)=0,\label{eq:EL-eq.}
\end{equation}
where the Euler operator $E_{A}^{\mu}$ of $A_{\mu}$ is defined by
\begin{equation}
E_{A}^{\mu}\equiv\sum_{i=1}^{n+1}\left(-1\right)^{i}D_{\mu_{1}}\cdots D_{\mu_{i}}\frac{\partial}{\partial\left(\partial_{\mu_{1}}\cdots\partial_{\mu_{i}}A_{\mu}\right)}.\label{eq:Euler-operator-A_=00005Cmu}
\end{equation}
Note that the Lagrangian density $\mathcal{L}_{F}$ depends on derivatives
of $A$ with respect to the spacetime coordinates up to the $(n+1)$-th
order. It includes the standard Maxwell system, i.e., 
\begin{equation}
\mathcal{L}_{F}=-\frac{1}{16\pi}F_{\mu\nu}F^{\mu\nu},\label{eq:Maxwell-Lagrangian}
\end{equation}
as a special case, where
\begin{equation}
F_{\mu\nu}=\partial_{\mu}A_{\nu}-\partial_{\nu}A_{\mu}.\label{eq:F=00005Cmu=00005Cnu}
\end{equation}
is the Faraday tensor. In Eq.\,(\ref{eq:Maxwell-Lagrangian}), $\mathcal{L}_{F}$
depends only on first-order derivatives of $A$. High-order electromagnetic
field theories appear in the study of gyrokinetic systems \citep{Qin2005,Qin2007a,Fan2020}
for magnetized plasmas and radiation reaction for classical charged
particles \citep{Bopp1940,Podolsky1942}. Physics requires that the
EL equation (\ref{eq:EL-eq.}) is gauge symmetric, i.e, invariant
under the gauge transformation $A_{\mu}\mapsto A_{\mu}+\partial_{\mu}f$.
In the present study, we assume that $\mathcal{L}_{F}$ is explicitly
gauge symmetric in the form of 
\begin{equation}
\mathcal{L}_{F}=\mathcal{L}_{F}\left(x^{\mu},F_{\mu\nu},DF_{\mu\nu},\cdots,D^{\left(n\right)}F_{\mu\nu}\right).\label{eq:cov-EM-Lagrangian}
\end{equation}
From the variational principle, $\delta\mathcal{A}=\delta\int\mathcal{L}_{F}d^{4}x=0$,
we have 
\begin{align}
 & 0=\delta\int\mathcal{L}_{F}d^{4}x=\int\left\{ E_{F}^{\mu\nu}\left(\mathcal{L}_{F}\right)\delta F_{\mu\nu}\right\} d^{4}x=\int\left\{ E_{F}^{\mu\nu}\left(\mathcal{L}_{F}\right)\left(\partial_{\mu}\delta A_{\nu}-\partial_{\nu}\delta A_{\mu}\right)\right\} d^{4}x\nonumber \\
 & =-\int\partial_{\mu}\left\{ 2E_{F}^{\left[\mu\nu\right]}\left(\mathcal{L}_{F}\right)\right\} \delta A_{\nu}d^{4}x,\label{eq:cov-Hamil-principle}
\end{align}
where the boundary term has been dropped, and $E_{F}^{\mu\nu}$ denotes
the Euler operator for the Faraday tensor $F_{\mu\nu}$ defined by
\begin{equation}
E_{F}^{\mu\nu}\left(\mathcal{L}_{F}\right)=\frac{\partial\mathcal{L}_{F}}{\partial F_{\mu\nu}}+\sum_{i=1}^{n}\left(-1\right)^{i}D_{\mu_{1}}\cdots D_{\mu_{i}}\frac{\partial\mathcal{L}_{F}}{\partial\partial_{\mu_{1}}\cdots\partial_{\mu_{i}}F_{\mu\nu}}\,.\label{eq:cov-Euler-ope}
\end{equation}
In Eq.\,(\ref{eq:cov-Hamil-principle}), superscript $\left[\mu\nu\right]$
represents anti-symmetrization with respect to $\mu$ and $\nu$,
i.e., 
\begin{equation}
E_{F}^{\left[\mu\nu\right]}\left(\mathcal{L}_{F}\right)\equiv\frac{1}{2}\left[E_{F}^{\mu\nu}\left(\mathcal{L}_{F}\right)-E_{F}^{\nu\mu}\left(\mathcal{L}_{F}\right)\right].\label{eq:anti-symmetrization}
\end{equation}
Due to the arbitrariness of $\delta A_{\nu}$ in Eq.\,(\ref{eq:cov-Hamil-principle}),
the equation of motion for the system is
\begin{equation}
\partial_{\mu}\mathcal{D}^{\mu\nu}=0,\label{eq:cov-EL-eq}
\end{equation}
where
\begin{equation}
\mathcal{D}^{\mu\nu}\equiv-8\pi E_{F}^{\left[\mu\nu\right]}\left(\mathcal{L}_{F}\right)\label{eq:electric-displacement-tensor}
\end{equation}
is the electric displacement tensor. 

In Sec.\,\ref{sec:Particle-Field-system}, we will consider electromagnetic
systems coupled with charged particles, and the Lagrangian density
$\mathcal{L}$ will depend on 4-potential $A_{\mu}$ i.e., 
\[
\mathcal{L}=\mathcal{L}\left(x^{\mu},A_{\mu},F_{\mu\nu},\cdots,D^{\left(n\right)}F_{\mu\nu}\right).
\]
Equation (\ref{eq:cov-EL-eq}) then becomes
\begin{equation}
\partial_{\mu}\mathcal{D}^{\mu\nu}=\frac{4\pi}{c}J_{f}^{\nu},\label{eq:Maxwell-eq-source}
\end{equation}
where 
\begin{equation}
J_{f}^{\nu}\equiv-c\frac{\partial\mathcal{L}_{F}}{\partial A_{\nu}}\label{eq:4-current}
\end{equation}
is the free 4-current.

For the standard Maxwell system (\ref{eq:Maxwell-Lagrangian}) without
free 4-current, the electric displacement tensor is the Faraday tensor,
i.e., $\mathcal{D}^{\mu\nu}=F^{\mu\nu}$, and Eq.\,(\ref{eq:cov-EL-eq})
reduces to Maxwell's equation
\begin{equation}
\partial_{\mu}F^{\mu\nu}=0.\label{eq:Maxwell-Eq.}
\end{equation}

\subsection{Infinitesimal criterion of symmetry and conservation laws \label{subsec:cov-Infinitesimal-criterion}}

A continuous symmetry of the action $\mathcal{A}$ is a group of transformation
\begin{equation}
\left(x^{\mu},A^{\nu}\right)\mapsto\left(\tilde{x}^{\mu},\tilde{A}^{\nu}\right)=g_{\epsilon}\cdot\left(x^{\mu},A^{\nu}\right),\label{eq:covr-transformation}
\end{equation}
such that 
\begin{equation}
\int\mathcal{L}_{F}\left(\tilde{x}^{\mu},\tilde{F}_{\mu\nu},\tilde{D}\tilde{F}_{\mu\nu},\cdots,\tilde{D}^{\left(n\right)}\tilde{F}_{\mu\nu}\right)d^{4}\tilde{x}=\int\mathcal{L}_{F}\left(x^{\mu},F_{\mu\nu},DF_{\mu\nu},\cdots,D^{\left(n\right)}F_{\mu\nu}\right)d^{4}x,\label{eq:cov-symmetry-condi}
\end{equation}
where $g_{\epsilon}$ constitutes a continuous group of the transformations
parameterized by $\epsilon$ \citep{Olver1993}. The infinitesimal
generator of the transformation group is
\begin{equation}
\boldsymbol{v}\coloneqq\frac{d}{d\epsilon}|_{0}g_{\epsilon}\cdot\left(x^{\mu},A^{\nu}\right)=\xi^{\mu}\frac{\partial}{\partial x^{\mu}}+\phi_{\mu}\frac{\partial}{\partial A_{\mu}}.\label{eq:cov-infinit-generator}
\end{equation}
By rewriting the symmetry condition (\ref{eq:cov-symmetry-condi})
as
\begin{equation}
\frac{d}{d\epsilon}|_{0}\int\mathcal{L}_{F}\left(\tilde{x}^{\mu},\tilde{F}_{\mu\nu},\tilde{D}\tilde{F}_{\mu\nu},\cdots,\tilde{D}^{\left(n\right)}\tilde{F}_{\mu\nu}\right)d^{4}\tilde{x}=0,\label{eq:cov-d/d=00005Cepsilon-1}
\end{equation}
we can derive the following infinitesimal version of the symmetry
condition,
\begin{equation}
\text{pr}^{\left(n+1\right)}\boldsymbol{v}\left(\mathcal{L}_{F}\right)+\mathcal{L}_{F}D_{\mu}\xi^{\mu}=0,\label{eq:cov-infinitesimal-criterion}
\end{equation}
where $\text{pr}^{\left(n+1\right)}\boldsymbol{v}$ is the prolongation
of $\boldsymbol{v}$. The standard prolongation formula for $\text{pr}^{\left(n+1\right)}\boldsymbol{v}$
can be found in Ref.$\thinspace$\citep{Olver1993}. In the present
study, we rewrite the prolongation formula with respect to $F_{\mu\nu}$,
instead of $A_{\mu}$, as
\begin{align}
 & \text{pr}^{\left(n+1\right)}\boldsymbol{v}=\boldsymbol{v}+\left[G_{\sigma\rho}+\xi^{\sigma}D_{\sigma}F_{\sigma\rho}\right]\frac{\partial\mathcal{L}_{F}}{\partial F_{\sigma\rho}}\nonumber \\
 & +\sum_{i=1}^{n}\left[D_{\mu_{1}}\cdots D_{\mu_{i}}G_{\sigma\rho}+\xi^{\sigma}D_{\sigma}D_{\mu_{1}}\cdots D_{\mu_{i}}F_{\sigma\rho}\right]\frac{\partial\mathcal{L}_{F}}{\partial\left(\partial_{\mu_{1}}\cdots\partial_{\mu_{i}}F_{\sigma\rho}\right)},\label{eq:cov-prolongation}
\end{align}
where 
\begin{equation}
G_{\sigma\rho}=\partial_{\sigma}Q_{\rho}-\partial_{\rho}Q_{\sigma}\equiv2\partial_{[\sigma}Q_{\rho]}\label{eq:cov-G-anti-cha}
\end{equation}
and 
\begin{equation}
Q_{\nu}=\phi_{\nu}-\xi^{\sigma}D_{\sigma}A_{\nu},\label{eq:cov-cha}
\end{equation}
is a characteristic of the Lie algebra.

Combining Eqs.\,(\ref{eq:cov-EL-eq}) and (\ref{eq:cov-infinitesimal-criterion})
generates the conservation law corresponding to the symmetry, 
\begin{equation}
D_{\mu}\left\{ \mathcal{L}_{F}\xi^{\mu}-\frac{1}{4\pi}\mathcal{D}^{\mu\nu}\left(\mathcal{L}_{F}\right)Q_{\nu}+\mathbb{P}^{\mu}\right\} =0,\label{eq:cov-conser}
\end{equation}
where
\begin{equation}
\mathbb{P}^{\mu}=\sum_{i=1}^{n}\sum_{j=1}^{i}\left(-1\right)^{j+1}\left(D_{\mu_{j+1}}\cdots D_{\mu_{i}}G_{\sigma\rho}\right)\left[D_{\mu_{1}}\cdots D_{\mu_{j-1}}\frac{\partial\mathcal{L}_{F}}{\partial\left(\partial_{\mu_{1}}\cdots\partial_{\mu_{j-1}}\partial_{\mu}\partial_{\mu_{j+1}}\cdots\partial_{\mu_{i}}F_{\sigma\rho}\right)}\right].\label{eq:cov-P}
\end{equation}
The conservation law given by Eq.\,(\ref{eq:cov-conser}) is not
gauge-symmetric in general.

\subsection{Gauge-symmetrization of the canonical EMT}

Now we assume the high-order electromagnetic field theory admits the
spacetime translation symmetry, i.e.,
\begin{equation}
\frac{\partial\mathcal{L}_{F}}{\partial x^{\mu}}=0,\label{eq:translation-symmetry}
\end{equation}
and derive the corresponding energy-momentum conservation law. Because
of Eq.\,(\ref{eq:translation-symmetry}), the action is invariant
under the spacetime translation 
\begin{equation}
\left(x^{\mu},A_{\nu}\right)\mapsto\left(x^{\mu}+\epsilon X_{0}^{\mu},A_{\nu}\right),\label{eq:space-time-translation}
\end{equation}
where $X_{0}^{\mu}$ is 4D constant vector field. The infinitesimal
generator $\boldsymbol{v}$, characteristic $Q^{\nu}$, and $G_{\sigma\rho}$
in Eq.\,(\ref{eq:cov-G-anti-cha}) are 
\begin{align}
 & \boldsymbol{v}=X_{0}^{\mu}\frac{\partial}{\partial x^{\mu}},\label{eq:cov-inf-gen-space-time}\\
 & Q^{\nu}=-X_{0}^{\nu}\partial_{\nu}A_{\sigma},\label{eq:cov-cha-space-time}\\
 & G_{\sigma\rho}=-X_{0}^{\nu}\partial_{\nu}F_{\sigma\rho}.\label{eq:cov-cha-anti-space-time}
\end{align}
The Lagrangian density satisfies the infinitesimal criterion because
\begin{equation}
X_{0}^{\mu}\frac{\partial\mathcal{L}_{F}}{\partial x^{\mu}}=0,\label{eq:cov-inf-criterion}
\end{equation}
which implies a conservation law. Substituting Eqs.\,(\ref{eq:cov-inf-gen-space-time})-(\ref{eq:cov-cha-anti-space-time})
into to Eq.\,(\ref{eq:cov-conser}), we obtain the canonical energy-momentum
conservation law according the standard Noether procedure,
\begin{align}
 & D_{\mu}T_{\text{N}}^{\mu\nu}=0,\label{eq:cov-cano-energy-momentum-conser}\\
 & T_{\text{N}}^{\mu\nu}=\mathcal{L}_{F}\eta^{\mu\nu}+\frac{1}{4\pi}\mathcal{D}^{\mu\sigma}\partial^{\nu}A_{\sigma}-\Sigma^{\mu\nu},\label{eq:canonical-EMT}\\
 & \Sigma^{\mu\nu}=\sum_{i=1}^{n}\sum_{j=1}^{i}\left(-1\right)^{j+1}\left(D_{\mu_{j+1}}\cdots D_{\mu_{i}}\partial^{\nu}F_{\sigma\rho}\right)\left[D_{\mu_{1}}\cdots D_{\mu_{j-1}}\frac{\partial\mathcal{L}_{F}}{\partial\left(\partial_{\mu_{1}}\cdots\partial_{\mu_{j-1}}\partial_{\mu}\partial_{\mu_{j+1}}\cdots\partial_{\mu_{i}}F_{\sigma\rho}\right)}\right].
\end{align}
In Eq.\,(\ref{eq:cov-cano-energy-momentum-conser}), $T_{\text{N}}^{\mu\nu}$
is the canonical EMT derived from the standard Noether procedure.

Obviously, $T_{\text{N}}^{\mu\nu}$ depends on the gauge as expected.
In the expression of $T_{\text{N}}^{\mu\nu}$ given by Eq.\,(\ref{eq:canonical-EMT}),
the gauge dependence comes from the second term, and the first and
third terms are gauge symmetric. Now we show how to gauge-symmetrize
$T_{\text{N}}^{\mu\nu}$. Note that because electric displacement
tensor $\mathcal{D}^{\sigma\mu}$ is anti-symmetric, the following
equations hold,
\begin{align}
 & D_{\mu}\left(D_{\sigma}\mathcal{F}^{\sigma\mu\nu}\right)=0,\label{eq:cov-identity}\\
 & \mathcal{F}^{\sigma\mu\nu}\equiv\frac{1}{4\pi}\mathcal{D}^{\sigma\mu}A^{\nu}.
\end{align}
Here, $\mathcal{F}^{\sigma\mu\nu}$ is a super-potential that is anti-symmetric
with respect to the first two indices. For easy reference, we will
call $\mathcal{F}^{\sigma\mu\nu}$ displacement-potential tensor.
The divergence of $\mathcal{F}^{\sigma\mu\nu}$ defines a divergence-free
tensor, i.e., 
\begin{equation}
T_{0}^{\mu\nu}\equiv D_{\sigma}\mathcal{F}^{\sigma\mu\nu}=-\frac{1}{4\pi}\mathcal{D}^{\mu\sigma}\partial_{\sigma}A^{\nu},\label{eq:cov-identity-1}
\end{equation}
where used is made of Eq.\,(\ref{eq:cov-EL-eq}). When $T_{0}^{\mu\nu}$
is added to $T_{\text{N}}^{\mu\nu},$ the gauge dependence is removed,
i.e., 
\begin{align}
 & D_{\mu}T_{\text{GS}}^{\mu\nu}=0,\label{eq:gauge-inv-EM-conser}\\
 & T_{\text{GS}}^{\mu\nu}\equiv T_{\text{N}}^{\mu\nu}+T_{0}^{\mu\nu}=\mathcal{L}_{F}\eta^{\mu\nu}+\frac{1}{4\pi}\mathcal{D}^{\mu\sigma}F_{\;\sigma}^{\nu}-\Sigma^{\mu\nu},\label{eq:gauge-inv-EMT}
\end{align}
where $T_{\text{GS}}^{\mu\nu}$ is the gauge-symmetric EMT.

It is worthwhile to mention that we derived the gauge-symmetrized
EMT $T_{\text{GS}}^{\mu\nu}$ from the expression of $T_{\text{N}}^{\mu\nu}$
in Eq.\,(\ref{eq:canonical-EMT}), which is calculated from the prolongation
with respect to $F_{\mu\nu}$. On the other hand, had we started from
Eq.\,(\ref{eq:cov-EM-Lagrangian-A_=00005Cmu}) and calculated the
EMT from the prolongation with respect to $A_{\mu}$, we would have
obtained a canonical EMT in the form of 
\begin{equation}
T_{\text{N}}^{\mu\nu}=\mathcal{L}_{F}\eta^{\mu\nu}-\hat{\Sigma}^{\mu\nu},\label{eq:canonical-A}
\end{equation}
where
\begin{equation}
\hat{\Sigma}^{\mu\nu}=\sum_{i=1}^{n+1}\sum_{j=1}^{i}\left(-1\right)^{j+1}D_{\mu_{j+1}}\cdots D_{\mu_{i}}\left(\partial_{\nu}A_{\sigma}\right)\left[D_{\mu_{1}}\cdots D_{\mu_{j-1}}\frac{\partial\mathcal{L}_{F}}{\partial\left(\partial_{\mu_{1}}\cdots\partial_{\mu_{j-1}}\partial_{\mu}\partial_{\mu_{j+1}}\cdots\partial_{\mu_{i}}A_{\sigma}\right)}\right].\label{eq:hat-sigma}
\end{equation}
However, different from the situation in Eq.\,(\ref{eq:canonical-EMT}),
every term in Eq.\,(\ref{eq:hat-sigma}) is gauge dependent, making
the gauge symmetrization difficult, if not impossible.

For the standard Maxwell electromagnetic system specified by Eq.\,(\ref{eq:Maxwell-Lagrangian}),
the electric displacement tensor reduces to the Faraday tensor $F^{\sigma\mu}$,
and the displacement-potential tensor $\mathcal{F}^{\sigma\mu\nu}$
reduces to $F^{\sigma\mu}A^{\nu}/4\pi$, coinciding with the tensor
used by Blaschke et al. for the $\text{U\ensuremath{\left(1\right)}}$
gauge theory \citep{Blaschke2016}.

\subsection{Comparison with the BR method}

As described above, the method proposed in the present study employs
displacement-potential tensor $\mathcal{F}^{\sigma\mu\nu}$ to gauge-symmetrize
the EMT, while the BR method use the super-potential $\mathcal{S}^{\sigma\mu\nu}$
to symmetrize the EMT. In this subsection, we discuss the difference
between the displacement-potential tensor $\mathcal{F}^{\sigma\mu\nu}$
in Eq.\,(\ref{eq:cov-identity}) and the BR super-potential $\mathcal{S}^{\sigma\mu\nu}$.
To calculate $\mathcal{S}^{\sigma\mu\nu}$, we need to first derive
the 4D angular momentum conservation laws generated by the Lorentz
symmetry. Assume that system is invariant under rotational transformation
in 4D spacetime
\begin{equation}
\left(x^{\mu},A^{\nu}\right)\mapsto\left(\tilde{x}^{\mu},\tilde{A}^{\nu}\right)=\left(\Lambda_{\epsilon}^{\mu\sigma}x_{\sigma},\Lambda_{\epsilon}^{\nu\sigma}A_{\sigma}\right),\label{eq:cov-rotation-trans}
\end{equation}
where $\left\{ \Lambda_{\epsilon}^{\mu\sigma}\right\} $ is one-parameter
subgroup of the Lorentz group. The infinitesimal generator $\boldsymbol{v}$,
the characteristic $Q_{\rho}$, and the term $G_{s\rho}$ are calculated
respectively by Eqs.\,(\ref{eq:cov-infinit-generator}), (\ref{eq:cov-cha}),
and (\ref{eq:cov-G-anti-cha}) as
\begin{align}
 & \boldsymbol{v}=\frac{d}{d\epsilon}|_{0}\left(\Lambda_{\epsilon}^{\mu\sigma}x_{\sigma},\Lambda_{\epsilon}^{\nu\sigma}A_{\sigma}\right)=\left(\Omega^{\mu\sigma}x_{\sigma},\Omega^{\mu\sigma}A_{\sigma}\right),\label{eq:cov-rot-inf-gen}\\
 & Q_{\rho}=\phi_{\rho}-\xi^{\alpha}D_{\alpha}A_{\rho}=\Omega_{\rho\alpha}A^{\alpha}-\Omega_{\alpha\beta}x^{\beta}D^{\alpha}A_{\rho},\label{eq:cov-cha-rot}\\
 & G_{s\rho}=\Omega_{\rho\alpha}F_{s}^{\;\alpha}-\Omega_{s\alpha}F_{\rho}^{\;\alpha}-\Omega_{\alpha\beta}x^{\beta}\partial^{\alpha}F_{s\rho},\label{eq:G-anti-rot}
\end{align}
where the anti-symmetric tensor $\Omega^{\mu\sigma}=\left[d\Lambda_{\epsilon}^{\mu\sigma}/d\epsilon\right]_{0}$
is the Lie algebra element of the Lorentz group. Substituting Eqs.\,(\ref{eq:cov-rot-inf-gen})-(\ref{eq:G-anti-rot})
into Eq.\,(\ref{eq:cov-conser}), we obtain the angular momentum
conservation law in 4D spacetime, 
\begin{equation}
\Omega_{\nu\sigma}D_{\mu}\left\{ x^{\sigma}T_{\text{N}}^{\mu\nu}+2E_{F}^{\left[\mu\nu\right]}\left(\mathcal{L}_{F}\right)A^{\sigma}+L^{\mu\nu\sigma}\right\} =0,
\end{equation}
which can be rewritten as 
\begin{equation}
D_{\mu}\left\{ \left[x^{\sigma}T_{\text{N}}^{\mu\nu}-x^{\nu}T_{\text{N}}^{\mu\sigma}\right]+S^{\mu\nu\sigma}\right\} =0.\label{eq:cov-angular-momen}
\end{equation}
In above equations,
\begin{align}
 & L^{\mu\nu\sigma}=\sum_{i=1}^{n}\sum_{j=1}^{i}\left(-1\right)^{j+1}D_{\mu_{j+1}}\cdots D_{\mu_{i}}\left(F_{\rho}^{\;\sigma}\right)D_{\mu_{1}}\cdots D_{\mu_{j-1}}\times\nonumber \\
 & \left[\frac{\partial\mathcal{L}}{\partial\left(\partial_{\mu_{1}}\cdots\partial_{\mu_{j-1}}\partial_{\mu}\partial_{\mu_{j+1}}\cdots\partial_{\mu_{i}}F_{\rho\nu}\right)}-\frac{\partial\mathcal{L}}{\partial\left(\partial_{\mu_{1}}\cdots\partial_{\mu_{j-1}}\partial_{\mu}\partial_{\mu_{j+1}}\cdots\partial_{\mu_{i}}F_{\nu\rho}\right)}\right]\nonumber \\
 & -\left[\sum_{i=1}^{n}\sum_{j=1}^{i}\left(-1\right)^{j+1}D_{\mu_{j+1}}\cdots D_{\mu_{i}}\left(x^{\sigma}\partial^{\nu}F_{s\rho}\right)D_{\mu_{1}}\cdots D_{\mu_{j-1}}\frac{\partial\mathcal{L}}{\partial\left(\partial_{\mu_{1}}\cdots\partial_{\mu_{j-1}}\partial_{\mu}\partial_{\mu_{j+1}}\cdots\partial_{\mu_{i}}F_{s\rho}\right)}-x^{\sigma}\Sigma^{\mu\nu}\right]\label{eq:cov-L}\\
 & \frac{1}{2}S^{\mu\nu\sigma}=E_{F}^{\left[\mu\nu\right]}\left(\mathcal{L}_{F}\right)A^{\sigma}-E_{F}^{\left[\mu\sigma\right]}\left(\mathcal{L}_{F}\right)A^{\nu}+\Delta^{\mu\nu\sigma},\label{eq:cov-s=00005Cmu=00005Cnu=00005Csig}\\
 & \Delta^{\mu\nu\sigma}\equiv2L^{\mu\left[\nu\sigma\right]}\nonumber \\
 & =\sum_{i=1}^{n}\sum_{j=1}^{i}\left(-1\right)^{j+1}D_{\mu_{j+1}}\cdots D_{\mu_{i}}\left\{ \left(F_{\rho}^{\;\sigma}\right)D_{\mu_{1}}\cdots D_{\mu_{j-1}}\left[\frac{\partial\mathcal{L}}{\partial\left(\partial_{\mu_{1}}\cdots\partial_{\mu_{j-1}}\partial_{\mu}\partial_{\mu_{j+1}}\cdots\partial_{\mu_{i}}F_{\rho\nu}\right)}\right.\right.\nonumber \\
 & \left.\vphantom{\left(F_{\rho}^{\;\sigma}\right)}-\frac{\partial\mathcal{L}}{\partial\left(\partial_{\mu_{1}}\cdots\partial_{\mu_{j-1}}\partial_{\mu}\partial_{\mu_{j+1}}\cdots\partial_{\mu_{i}}F_{\nu\rho}\right)}\right]-\left(F_{\rho}^{\;\nu}\right)D_{\mu_{1}}\cdots D_{\mu_{j-1}}\nonumber \\
 & \left.\vphantom{\left(F_{\rho}^{\;\sigma}\right)}\times\left[\frac{\partial\mathcal{L}}{\partial\left(\partial_{\mu_{1}}\cdots\partial_{\mu_{j-1}}\partial_{\mu}\partial_{\mu_{j+1}}\cdots\partial_{\mu_{i}}F_{\rho\sigma}\right)}-\frac{\partial\mathcal{L}}{\partial\left(\partial_{\mu_{1}}\cdots\partial_{\mu_{j-1}}\partial_{\mu}\partial_{\mu_{j+1}}\cdots\partial_{\mu_{i}}F_{\sigma\rho}\right)}\right]\right\} \nonumber \\
 & -\sum_{i=1}^{n}\sum_{j=1}^{i}\left(-1\right)^{j+1}D_{\mu_{j+1}}\cdots D_{\mu_{i}}\left(x^{\sigma}\partial^{\nu}F_{s\rho}-x^{\nu}\partial^{\sigma}F_{s\rho}\right)D_{\mu_{1}}\cdots D_{\mu_{j-1}}\times\nonumber \\
 & \frac{\partial\mathcal{L}}{\partial\left(\partial_{\mu_{1}}\cdots\partial_{\mu_{j-1}}\partial_{\mu}\partial_{\mu_{j+1}}\cdots\partial_{\mu_{i}}F_{s\rho}\right)}+x^{\sigma}\Sigma^{\mu\nu}-x^{\nu}\Sigma^{\mu\sigma},\label{eq:cov-Delta}
\end{align}
where the superscript $\left[\mu\nu\right]$ denotes anti-symmetrization
with respect to $\mu$ and $\nu$.

The anti-symmetric BR super-potentail $\mathcal{S}^{\sigma\mu\nu}$
is defined from the tensor $S^{\sigma\mu\nu}$ in Eq.\,(\ref{eq:cov-angular-momen})
as \citep{Belinfante1939,Belinfante1940,Rosenfeld1940}
\begin{align}
\mathcal{S}^{\sigma\mu\nu} & \equiv\frac{1}{2}\left[S^{\sigma\nu\mu}-S^{\mu\nu\sigma}-S^{\nu\mu\sigma}\right].\label{eq:cov-BR-tensor}
\end{align}
It is clear from Eqs.\,(\ref{eq:cov-L})-(\ref{eq:cov-Delta}) that
$\mathcal{S}^{\sigma\mu\nu}$ and $\mathcal{F}^{\sigma\mu\nu}$ are
related as follows,

\begin{equation}
\mathcal{S}^{\sigma\mu\nu}=\mathcal{F}^{\sigma\mu\nu}+\frac{1}{2}\left[\Delta^{\sigma\nu\mu}-\Delta^{\mu\nu\sigma}-\Delta^{\nu\mu\sigma}\right].
\end{equation}
Equation (\ref{eq:cov-BR-tensor}) shows that in general $\mathcal{F}^{\sigma\mu\nu}$
is different from $\mathcal{S}^{\sigma\mu\nu}$ when $\Delta^{\sigma\nu\mu}$
is non-vanishing. For a first-order field theory, such as the standard
Maxwell system (\ref{eq:Maxwell-Lagrangian}), $n=1$ and the last
three terms vanish such that $\mathcal{S}^{\sigma\mu\nu}=\mathcal{F}^{\sigma\mu\nu}$.
In this situation, adding $T_{0}^{\mu\nu}\equiv D_{\sigma}\mathcal{F}^{\sigma\mu\nu}$
to $T_{\text{N}}^{\mu\nu}$ will render it both symmetric and gauge-symmetric,
and the method developed here can be used as a simpler procedure to
calculate the BR super-potential $\mathcal{S}^{\sigma\mu\nu}$ without
the necessity to calculate the angular momentum tensor in 4D spacetime. 

As an example of high-order electromagnetic field theory, we consider
the Podolsky system \citep{Bopp1940,Podolsky1942}, which was proposed
to study the radiation reaction of classical charged particles. The
Podolsky Lagrangian density is
\begin{equation}
\mathcal{L}_{\text{Po}}=\frac{1}{8\pi}\left\{ \boldsymbol{E}^{2}-\boldsymbol{B}^{2}+a^{2}\left[\left(\boldsymbol{\nabla}\cdot\boldsymbol{E}\right)^{2}-\left(\boldsymbol{\nabla}\times\boldsymbol{B}-\frac{1}{c}\partial_{t}\boldsymbol{E}\right)^{2}\right]\right\} \label{eq:BP-lagrangian}
\end{equation}
or in a manifestly covariant form 
\begin{equation}
\mathcal{L}_{\text{Po}}=-\frac{1}{16\pi}F_{\sigma\rho}F^{\sigma\rho}-\frac{a^{2}}{8\pi}\partial_{\sigma}F^{\sigma\lambda}\partial^{\rho}F_{\rho\lambda}.\label{eq:cov-BP-Lagrangian}
\end{equation}
We substitute the Lagrangian density (\ref{eq:cov-BP-Lagrangian})
into Eq.\,(\ref{eq:canonical-EMT}) to obtain the canonical EMT
\begin{align}
 & 4\pi T_{\text{N}}^{\mu\nu}=\left(-\frac{1}{4}F_{\sigma\rho}F^{\sigma\rho}-\frac{a^{2}}{2}\partial_{\sigma}F^{\sigma\lambda}\partial^{\rho}F_{\rho\lambda}\right)\eta^{\mu\nu}\nonumber \\
 & +\left[F^{\mu\sigma}-a^{2}\left(\partial^{\mu}\partial_{\lambda}F^{\lambda\sigma}-\partial^{\sigma}\partial_{\lambda}F^{\lambda\mu}\right)\right]\partial^{\nu}A_{\sigma}+a^{2}\left(\partial^{\nu}F_{\;\rho}^{\mu}\right)\left(\partial_{\sigma}F^{\sigma\rho}\right),\label{eq:Po-canonical-EM-conser}
\end{align}
where use is made of the following equations,
\begin{align}
 & \frac{\partial}{\partial\left(\partial_{\sigma}F_{\mu\nu}\right)}\left[\partial_{\alpha}F^{\alpha\lambda}\partial^{\rho}F_{\rho\lambda}\right]=2\eta^{\sigma\mu}\partial_{\lambda}F^{\lambda\nu},\\
 & D_{\sigma}\frac{\partial}{\partial\left(\partial_{\sigma}F_{\mu\nu}\right)}\left[\partial_{\alpha}F^{\alpha\lambda}\partial^{\rho}F_{\rho\lambda}\right]=2\partial^{\mu}\partial_{\sigma}F^{\sigma\nu},\\
 & E_{F}^{\mu\sigma}=\frac{\partial\mathcal{L}_{\text{Po}}}{\partial F_{\mu\sigma}}-D_{\rho}\frac{\partial\mathcal{L}_{\text{Po}}}{\partial\partial_{\rho}F_{\mu\sigma}}=-\frac{1}{8\pi}F^{\mu\sigma}+\frac{a^{2}}{4\pi}\partial^{\mu}\partial_{\lambda}F^{\lambda\sigma},\\
 & 2E_{F}^{\left[\mu\sigma\right]}=-\frac{1}{4\pi}F^{\mu\sigma}+\frac{a^{2}}{4\pi}\left(\partial^{\mu}\partial_{\lambda}F^{\lambda\sigma}-\partial^{\sigma}\partial_{\lambda}F^{\lambda\mu}\right),\\
 & \Sigma^{\mu\nu}=\left(\partial^{\nu}F_{\sigma\rho}\right)\frac{\partial\mathcal{L}_{\text{Po}}}{\partial\left(\partial_{\mu}F_{\sigma\rho}\right)}=-\frac{a^{2}}{4\pi}\partial^{\nu}F_{\sigma\rho}\left[\eta^{\mu\sigma}\partial_{\lambda}F^{\lambda\rho}\right]=-\frac{a^{2}}{4\pi}\left(\partial^{\nu}F_{\;\rho}^{\mu}\right)\left(\partial_{\sigma}F^{\sigma\rho}\right).
\end{align}
The displacement-potential tensor is
\begin{equation}
\mathcal{F}^{\mu\nu\sigma}\equiv\frac{1}{4\pi}\mathcal{D}^{\sigma\mu}A^{\nu}=\frac{1}{4\pi}\left[-F^{\mu\sigma}+a^{2}\left(\partial^{\mu}\partial_{\lambda}F^{\lambda\sigma}-\partial^{\sigma}\partial_{\lambda}F^{\lambda\mu}\right)\right]A^{\nu},\label{eq:Po-F-1}
\end{equation}
and
\begin{equation}
4\pi\partial_{\sigma}\mathcal{F}^{\mu\nu\sigma}=\left[-F^{\mu\sigma}+a^{2}\left(\partial^{\mu}\partial_{\lambda}F^{\lambda\sigma}-\partial^{\sigma}\partial_{\lambda}F^{\lambda\mu}\right)\right]\partial_{\sigma}A^{\nu}.\label{eq:Po-F-2}
\end{equation}
Adding Eq.\,(\ref{eq:Po-F-2}) to Eq.\,(\ref{eq:Po-canonical-EM-conser}),
we obtain the gauge-symmetric EMT,
\begin{align}
 & 4\pi T_{\text{GS}}^{\mu\nu}=\left[F^{\mu\sigma}F_{\nu\sigma}-\frac{1}{4}\left(F_{\sigma\rho}F^{\sigma\rho}\right)\eta^{\mu\nu}\right]-\frac{a^{2}}{2}\left(\partial_{\sigma}F^{\sigma\lambda}\partial^{\rho}F_{\rho\lambda}\right)\eta^{\mu\nu}\nonumber \\
 & -a^{2}F_{\;\sigma}^{\nu}\left(\partial^{\mu}\partial_{\rho}F^{\rho\sigma}\right)+a^{2}F_{\;\sigma}^{\nu}\left(\partial^{\sigma}\partial_{\rho}F^{\rho\mu}\right)+a^{2}\left(\partial^{\nu}F_{\;\rho}^{\mu}\right)\left(\partial_{\sigma}F^{\sigma\rho}\right).\label{eq:Po-EMT-inv}
\end{align}
It is easy to see that $T_{\text{GS}}^{\mu\nu}$ for the Podolsky
system is not symmetric, i.e., $T_{\text{GS}}^{\mu\nu}\neq T_{\text{GS}}^{\nu\mu}$.

To calculate the BR EMT $T_{\text{BR}}^{\mu\nu}$ for the Podolsky
system, we evaluate the $\Delta^{\mu\nu\sigma}$ term in Eq.\,(\ref{eq:cov-BR-tensor}).
Using Eq.\,(\ref{eq:cov-Delta}), we have
\begin{align}
\Delta^{\mu\nu\sigma} & =F_{\rho}^{\;\sigma}\left[\frac{\partial\mathcal{L}_{\text{Po}}}{\partial\left(\partial_{\mu}F_{\rho\nu}\right)}-\frac{\partial\mathcal{L}_{\text{Po}}}{\partial\left(\partial_{\mu}F_{\nu\rho}\right)}\right]-F_{\rho}^{\;\nu}\left[\frac{\partial\mathcal{L}_{\text{Po}}}{\partial\left(\partial_{\mu}F_{\rho\sigma}\right)}-\frac{\partial\mathcal{L}_{\text{Po}}}{\partial\left(\partial_{\mu}F_{\sigma\rho}\right)}\right]\nonumber \\
 & =-\frac{a^{2}}{4\pi}\left[F^{\mu\sigma}\partial_{\rho}F^{\rho\nu}-\eta^{\mu\nu}\left(F_{\rho}^{\;\sigma}\partial_{\lambda}F^{\lambda\rho}\right)-F^{\mu\nu}\partial_{\rho}F^{\rho\sigma}+\eta^{\mu\sigma}F_{\rho}^{\;\nu}\left(\partial_{\lambda}F^{\lambda\rho}\right)\right].\label{eq:Po-Delta}
\end{align}
Substituting Eqs.\,(\ref{eq:cov-BR-tensor}) and (\ref{eq:Po-Delta})
into Eq.\,(\ref{eq:T_BR}), we obtain the BR EMT as
\begin{align}
 & 4\pi T_{\text{BR}}^{\mu\nu}=4\pi T_{\text{GS}}^{\mu\nu}+2\pi\left[\Delta^{\sigma\nu\mu}-\Delta^{\mu\nu\sigma}-\Delta^{\nu\mu\sigma}\right]\nonumber \\
 & =\left[F^{\mu\sigma}F_{\nu\sigma}-\frac{1}{4}\left(F_{\sigma\rho}F^{\sigma\rho}\right)\eta^{\mu\nu}\right]+\frac{a^{2}}{2}\left[\left(\partial_{\sigma}F^{\sigma\rho}\right)\left(\partial^{\lambda}F_{\lambda\rho}\right)-2F_{\rho}^{\;\sigma}\left(\partial_{\sigma}\partial_{\lambda}F^{\lambda\rho}\right)\right]\eta^{\mu\nu}\nonumber \\
 & +a^{2}\left[F^{\nu\sigma}\left(\partial_{\sigma}\partial_{\rho}F^{\rho\mu}\right)+F^{\mu\sigma}\left(\partial_{\sigma}\partial_{\rho}F^{\rho\nu}\right)-\left(\partial_{\sigma}F^{\sigma\mu}\right)\left(\partial_{\rho}F^{\rho\nu}\right)\right.\nonumber \\
 & \left.\vphantom{\left(\partial_{\rho}F^{\rho\nu}\right)}-F_{\;\sigma}^{\nu}\left(\partial^{\mu}\partial_{\rho}F^{\rho\sigma}\right)-F_{\;\sigma}^{\mu}\left(\partial^{\nu}\partial_{\rho}F^{\rho\sigma}\right)\right].\label{eq:Po-T_BR}
\end{align}
It is easy to verify that $T_{\text{BR}}^{\mu\nu}$ for the Podolsky
system is both symmetric and gauge-symmetric.

\section{Gauge-symmetric EMTs for electromagnetic systems coupled with classical
charged particles \label{sec:Particle-Field-system}}

For self-consistent electromagnetic systems with free currents, the
electromagnetic fields are coupled with charged particles. In this
section, we apply the theory established in Sec.\,\ref{sec:Gauge invariant conservation laws for electromagnetic system}
to derive gauge-symmetric EMTs for electromagnetic systems coupled
with classical charged particles. 

For practical applications, such as in the gyrokinetic theory \citep{Qin2005,Qin2007a,Fan2020}
for magnetized plasmas, reduced theoretical models are often adopted
due to the intrinsic complexity of the systems. The equations of motion
for the systems are usually gauge invariant, but the Lagrangian densities
are not always specified by manifestly covariant forms. For these
systems, energy and momentum conservation laws need to be derived
separately. We demonstrate how the energy and momentum conservation
laws can be transformed into gauge-symmetric forms using the ``3+1''
form of Eq.\,(\ref{eq:cov-identity}), i.e., 
\begin{equation}
\frac{D}{Dt}\left\{ \frac{D}{D\boldsymbol{x}}\cdot\left[\boldsymbol{E}_{\boldsymbol{E}}\left(\mathcal{L}\right)\varphi\right]\right\} +\frac{D}{D\boldsymbol{x}}\cdot\left\{ \frac{D}{Dt}\left[-\boldsymbol{E}_{\boldsymbol{E}}\left(\mathcal{L}\right)\varphi\right]\right\} =0\label{eq:PF-identity-1}
\end{equation}
and 
\begin{equation}
\frac{D}{Dt}\left\{ \frac{D}{D\boldsymbol{x}}\cdot\left[-\frac{1}{c}\boldsymbol{E}_{\boldsymbol{E}}\left(\mathcal{L}\right)\boldsymbol{A}\right]\right\} +\frac{D}{D\boldsymbol{x}}\cdot\left\{ \frac{D}{Dt}\left[\frac{1}{c}\boldsymbol{E}_{\boldsymbol{E}}\left(\mathcal{L}\right)\boldsymbol{A}\right]\right\} =0.\label{eq:PF-identity-2}
\end{equation}

\subsection{Weak Euler-Lagrange equation and conservation law \label{subsec:Weak-Euler-Lagrange-equation}}

The Lagrangian density of a generic classical electromagnetic field-charge
particle system assumes the form of 
\begin{align}
\mathcal{L} & =\sum_{a}\mathcal{L}_{a}+\mathcal{L}_{F},\label{eq:PF-Lagrangian-density-1}\\
\mathcal{L}_{a} & =L_{a}\delta_{a},\quad L_{a}=L_{a}\left(x^{\mu},\boldsymbol{X}_{a},\dot{\boldsymbol{X}}_{a};\varphi,\boldsymbol{A},\boldsymbol{E},\boldsymbol{B},D\boldsymbol{E},D\boldsymbol{B},\cdots,D^{n}\boldsymbol{E},D^{n}\boldsymbol{B}\right)\label{eq:particle-Lag-den}
\end{align}
where the subscript $a$ labels particles, $\boldsymbol{X}_{a}$ is
its trajectory, $\mathcal{L}_{a}$ is its Lagrangian density, and
$\delta_{a}\equiv\delta\left(\boldsymbol{x}-\boldsymbol{X}_{a}\right)$.
Here, $\delta(x)$ is the Dirac $\delta$-function.

In the ``3+1'' form, the equations of motion for the electromagnetic
field are
\begin{align}
 & \nabla\cdot\boldsymbol{E}_{\boldsymbol{E}}\left(\mathcal{L}\right)=-\frac{\partial\mathcal{L}}{\partial\varphi},\label{eq:PF-Maxwell's-Eq1}\\
 & -\frac{1}{c}\frac{\partial}{\partial t}\left[\boldsymbol{E}_{\boldsymbol{E}}\left(\mathcal{L}\right)\right]-\boldsymbol{\nabla}\times\left[\boldsymbol{E}_{\boldsymbol{B}}\left(\mathcal{L}\right)\right]=\frac{\partial\mathcal{L}}{\partial\boldsymbol{A}}.\label{eq:PF-Maxwell's-Eq2}
\end{align}
In this study, it is assumed that the Lagrangian density $\mathcal{L}$
is linear in terms of $\varphi$ and $\boldsymbol{A}$, and Eqs.\,(\ref{eq:PF-Maxwell's-Eq1})
and (\ref{eq:PF-Maxwell's-Eq2}) are thus gauge symmetric. Specifically,
we assume that $\mathcal{L}$ depends on $\varphi$ and $\boldsymbol{A}$
only through the term $-q_{a}\delta_{a}(\varphi+\boldsymbol{A}\cdot\dot{\boldsymbol{X}}_{a}/c)$,
i.e., the Lagrangian density can be written as
\begin{equation}
\mathcal{L}=\sum_{a}\left[-\varphi+\frac{1}{c}\boldsymbol{A}\cdot\dot{\boldsymbol{X}}_{a}\right]q_{a}\delta_{a}+\text{GSP}\left(\mathcal{L}\right),\label{eq:PF-Lag-den-2-gauge-seperated}
\end{equation}
where ``$\text{GSP}\left(\mathcal{L}\right)$'' denotes the gauge-symmetric
parts of the Lagrangian density $\mathcal{L}$. The right hand side
of Eqs.\,(\ref{eq:PF-Maxwell's-Eq1}) and (\ref{eq:PF-Maxwell's-Eq2})
are the ``3+1'' form of Eq.\,(\ref{eq:4-current}), the free charge
density $\rho_{f}$ and current density $\boldsymbol{j}_{f}$, respectively.
Using Eq.\,(\ref{eq:PF-Lag-den-2-gauge-seperated}), we have 
\begin{equation}
\rho_{f}=-\frac{\partial\mathcal{L}}{\partial\varphi}=\sum_{a}q_{a}\delta_{a},\quad\boldsymbol{j}_{f}=c\frac{\partial\mathcal{L}}{\partial\boldsymbol{A}}=\sum_{a}q_{a}\dot{\boldsymbol{X}}_{a}\delta_{a}.\label{eq:PF-charge-den-current}
\end{equation}

The equation of motion for particles is also derived from the variational
principle. However, because particles and field reside on different
manifolds, the equation of motion for particles will be the weak EL
equation \citep{Qin2014b,Fan2018,Fan2019,Fan2020}
\begin{equation}
\boldsymbol{E}_{\boldsymbol{X}_{a}}\left(\mathcal{L}\right)=\frac{D}{D\boldsymbol{x}}\cdot\left(\dot{\boldsymbol{X}}_{a}\frac{\partial\mathcal{L}}{\partial\dot{\boldsymbol{X}}_{a}}-\mathcal{L}_{a}\boldsymbol{I}\right),\label{eq:weak-EL}
\end{equation}
where $\boldsymbol{E}_{\boldsymbol{X}_{a}}$ is the Euler operator
for the trajectory of the $a$-th particle,
\begin{equation}
\boldsymbol{E}_{\boldsymbol{X}_{a}}=\frac{\partial}{\partial\boldsymbol{X}_{a}}-\frac{d}{dt}\frac{\partial}{\partial\dot{\boldsymbol{X}_{a}}}.\label{eq:Euler-operator-particle}
\end{equation}
To derive a local conservation law from a symmetry, we need the infinitesimal
symmetry criterion for the Lagrangian density. A symmetry of the action
$\mathcal{A}\equiv\int\mathcal{L}dtd^{3}\boldsymbol{x}$ is defined
by group transformations
\begin{equation}
\left(x^{\mu},\boldsymbol{X}_{a};\varphi,\boldsymbol{A}\right)\mapsto\left(\tilde{x}^{\mu},\tilde{\boldsymbol{X}}_{a};\tilde{\varphi},\tilde{\boldsymbol{A}}\right)=g_{\epsilon}\cdot\left(x^{\mu},\boldsymbol{X}_{a};\varphi,\boldsymbol{A}\right),\label{eq:PF-group-transform}
\end{equation}
such that 
\begin{equation}
\int\mathcal{L}\left(\tilde{x}^{\mu},\tilde{\boldsymbol{X}}_{a},\tilde{\boldsymbol{E}},\tilde{\boldsymbol{B}},\cdots,\tilde{D}^{n}\tilde{\boldsymbol{E}},\tilde{D}^{n}\tilde{\boldsymbol{B}}\right)d\tilde{t}d^{3}\tilde{\boldsymbol{x}}=\int\mathcal{L}\left(x^{\mu},\boldsymbol{X}_{a},\boldsymbol{E},\boldsymbol{B},\cdots,D^{\left(n\right)}\boldsymbol{E},D^{\left(n\right)}\boldsymbol{B}\right)dtd^{3}\boldsymbol{x}.
\end{equation}
The corresponding infinitesimal generator of (\ref{eq:PF-group-transform})
is 
\begin{equation}
\boldsymbol{v}=\xi^{\mu}\frac{\partial}{\partial x^{\mu}}+\sum_{a}\boldsymbol{\theta}_{a}\cdot\frac{\partial}{\partial\boldsymbol{X}_{a}}+\phi_{0}\frac{\partial}{\partial\varphi}+\boldsymbol{\phi}_{\boldsymbol{A}}\cdot\frac{\partial}{\partial\boldsymbol{A}}.\label{eq:PF-inf-generator}
\end{equation}
The infinitesimal criterion of the symmetry condition can be derived
using the same procedure in Sec.\,\ref{subsec:cov-Infinitesimal-criterion},
\begin{equation}
\text{pr}^{\left(n+1\right)}\boldsymbol{v}\left(\mathcal{L}\right)+\mathcal{L}D_{\mu}\xi^{\mu}=0.\label{eq:PF-infinitesimal-criterion}
\end{equation}
The prolongation of $\boldsymbol{v}$ now reads
\begin{align}
 & \text{pr}^{\left(n+1\right)}\boldsymbol{v}\left(\mathcal{L}\right)\nonumber \\
 & =\boldsymbol{v}+\sum_{a}\left[\left(\dot{\boldsymbol{q}}_{a}+\xi^{t}\ddot{\boldsymbol{X}}_{a}\right)\cdot\frac{\partial\mathcal{L}}{\partial\dot{\boldsymbol{X}}_{a}}\right]-\left[\boldsymbol{\nabla}Q_{0}+\xi^{\mu}D_{\mu}\left(\boldsymbol{\nabla}\varphi\right)\right]\cdot\frac{\partial\mathcal{L}}{\partial\boldsymbol{E}}\nonumber \\
 & -\cdots-\sum_{i=1}^{n}\left[D_{\mu_{1}}\cdots D_{\mu_{i}}\boldsymbol{\nabla}Q_{\alpha0}+\xi^{\mu}D_{\mu}D_{\mu_{1}}\cdots D_{\mu_{i}}\left(\boldsymbol{\nabla}\varphi_{\alpha}\right)\right]\cdot\frac{\partial\mathcal{L}}{\partial D_{\mu_{1}}\cdots D_{\mu_{i}}\boldsymbol{E}}\nonumber \\
 & -\left[D_{t}\boldsymbol{Q}_{\boldsymbol{A}}+\xi^{\mu}D_{\mu}\boldsymbol{A}_{,t}\right]\cdot\left(\frac{1}{c}\frac{\partial\mathcal{L}}{\partial\boldsymbol{E}}\right)\nonumber \\
 & -\cdots-\sum_{i=1}^{n}\left[D_{\mu_{1}}\cdots D_{\mu_{i}}D_{t}\boldsymbol{Q}_{\boldsymbol{A}}+\xi^{\mu}D_{\mu}D_{\mu_{1}}\cdots D_{\mu_{i}}\boldsymbol{A}_{,t}\right]\cdot\frac{\partial\mathcal{L}}{\partial D_{\mu_{1}}\cdots D_{\mu_{i}}\boldsymbol{E}}\nonumber \\
 & +\left[\boldsymbol{\nabla}\boldsymbol{Q}_{\boldsymbol{A}}+\xi^{\mu}D_{\mu}\boldsymbol{\nabla}\boldsymbol{A}\right]:\left(\boldsymbol{\varepsilon}\cdot\frac{\partial\mathcal{L}}{\partial\boldsymbol{B}}\right)\nonumber \\
 & +\cdots+\sum_{i=1}^{n}\left[D_{\mu_{1}}\cdots D_{\mu_{i}}\boldsymbol{\nabla}\boldsymbol{Q}_{\boldsymbol{A}}+\xi^{\mu}D_{\mu}D_{\mu_{1}}\cdots D_{\mu_{i}}\boldsymbol{\nabla}\boldsymbol{A}\right]:\left(\boldsymbol{\varepsilon}\cdot\frac{\partial\mathcal{L}}{\partial D_{\mu_{1}}\cdots D_{\mu_{i}}\boldsymbol{B}}\right),\label{eq:PF-prolongation}
\end{align}
where 
\begin{equation}
\boldsymbol{q}_{a}=\boldsymbol{\theta}_{a}-\xi^{t}\dot{\boldsymbol{X}}_{a}\label{eq:charact-particle}
\end{equation}
is another characteristic of $\boldsymbol{v}$ induced by particle's
trajectory. To obtain the corresponding conservation law, we transform
the infinitesimal criterion into 
\begin{align}
 & \partial_{t}\left[\mathcal{L}\xi^{t}-\frac{1}{c}\boldsymbol{Q}_{\boldsymbol{A}}\cdot\boldsymbol{E}_{\boldsymbol{E}}\left(\mathcal{L}\right)+\sum_{a}\left(\boldsymbol{q}_{a}\cdot\frac{\partial\mathcal{L}}{\partial\dot{\boldsymbol{X}}_{a}}\right)\right]+\boldsymbol{\nabla}\cdot\left[\mathcal{L}\boldsymbol{\kappa}-Q_{0}\boldsymbol{E}_{\boldsymbol{E}}\left(\mathcal{L}\right)+\boldsymbol{Q}_{\boldsymbol{A}}\times\boldsymbol{E}_{\boldsymbol{B}}\left(\mathcal{L}\right)\right]\nonumber \\
 & +D_{\mu}\left[\mathcal{\mathbb{P}}_{1}^{\mu}+\mathcal{\mathbb{P}}_{2}^{\mu}\right]+\sum_{a}\left[\boldsymbol{q}_{a}\cdot\boldsymbol{E}_{\boldsymbol{X}_{a}}\left(\mathcal{L}\right)\right]+\left\{ \frac{\partial\mathcal{L}}{\partial\varphi}+\boldsymbol{\nabla}\cdot\left[\boldsymbol{E}_{\boldsymbol{E}}\left(\mathcal{L}\right)\right]\right\} Q_{0}\nonumber \\
 & +\left\{ \frac{\partial\mathcal{L}}{\partial\boldsymbol{A}}+\frac{1}{c}\partial_{t}\left[\boldsymbol{E}_{\boldsymbol{E}}\left(\mathcal{L}\right)\right]+\left[\boldsymbol{\nabla}\times\boldsymbol{E}_{\boldsymbol{B}}\left(\mathcal{L}\right)\right]\right\} \cdot\boldsymbol{Q}_{\boldsymbol{A}}=0,\label{eq:PF-infinitesimal-trans}
\end{align}
where 
\begin{align}
\mathcal{\mathbb{P}}_{1}^{\mu} & =\sum_{i=1}^{n}\sum_{j=1}^{i}\left(-1\right)^{j}D_{\mu_{j+1}}\cdots D_{\mu_{i}}\left(\boldsymbol{\nabla}Q_{0}+\frac{1}{c}D_{t}\boldsymbol{Q}_{\boldsymbol{A}}\right)\nonumber \\
 & \cdot\left[D_{\mu_{1}}\cdots D_{\mu_{j-1}}\frac{\partial\mathcal{L}}{\partial D_{\mu_{1}}\cdots D_{\mu_{j-1}}D_{\mu}D_{\mu_{j+1}}\cdots D_{\mu_{i}}\boldsymbol{E}}\right],\label{eq:73}\\
\mathbb{P}_{2}^{\mu} & =\sum_{i=1}^{n}\sum_{j=1}^{i}\left(-1\right)^{j}D_{\mu_{j+1}}\cdots D_{\mu_{i}}\left(-\boldsymbol{\nabla}\times\boldsymbol{Q}_{\boldsymbol{A}}\right)\nonumber \\
 & \cdot\left[D_{\mu_{1}}\cdots D_{\mu_{j-1}}\frac{\partial\mathcal{L}}{\partial D_{\mu_{1}}\cdots D_{\mu_{j-1}}D_{\mu}D_{\mu_{j+1}}\cdots D_{\mu_{i}}\boldsymbol{B}}\right].\label{eq:74}
\end{align}
The last two terms on the left-hand-side of Eq.\,(\ref{eq:PF-infinitesimal-trans})
vanish due to Eqs.\,(\ref{eq:PF-Maxwell's-Eq1}) and (\ref{eq:PF-Maxwell's-Eq2}),
but the fourth term does not because of the weak EL equation (\ref{eq:weak-EL}).
If the characteristic $\boldsymbol{q}_{a}$ is independent of $\boldsymbol{x}$,
$\boldsymbol{E}$, and $\boldsymbol{B}$, the conservation law of
the symmetry is established as
\begin{align}
 & \partial_{t}\left[\mathcal{L}\xi^{t}-\frac{1}{c}\boldsymbol{Q}_{\boldsymbol{A}}\cdot\boldsymbol{E}_{\boldsymbol{E}}\left(\mathcal{L}\right)+\sum_{a}\left(\boldsymbol{q}_{a}\cdot\frac{\partial\mathcal{L}}{\partial\dot{\boldsymbol{X}}_{a}}\right)\right]+\boldsymbol{\nabla}\cdot\left[\mathcal{L}\boldsymbol{\kappa}-Q_{0}\boldsymbol{E}_{\boldsymbol{E}}\left(\mathcal{L}\right)\right.\nonumber \\
 & \left.\vphantom{\frac{\partial\mathcal{L}_{a}}{\partial\dot{\boldsymbol{X}}_{a}}}+\boldsymbol{Q}_{\boldsymbol{A}}\times\boldsymbol{E}_{\boldsymbol{B}}\left(\mathcal{L}\right)+\sum_{a}\left(\dot{\boldsymbol{X}}_{a}\frac{\partial\mathcal{L}_{a}}{\partial\dot{\boldsymbol{X}}_{a}}-\mathcal{L}_{a}\boldsymbol{I}\right)\cdot\boldsymbol{q}_{a}\right]+D_{\mu}\left[\mathcal{\mathbb{P}}_{1}^{\mu}+\mathcal{\mathbb{P}}_{2}^{\mu}\right]=0.\label{eq:PF-general-conservation}
\end{align}

\subsection{Gauge-symmetric energy conservation law \label{subsec:PF-The-improvement-energy}}

We first derive the gauge-symmetric energy conservation law, assuming
that the action $\mathcal{A}\equiv\int\mathcal{L}dtd^{3}\boldsymbol{x}$
is unchanged under the time translation
\begin{equation}
\left(t,\boldsymbol{x},\boldsymbol{X}_{a},\varphi,\boldsymbol{A}\right)\mapsto\left(t+\epsilon,\boldsymbol{x},\boldsymbol{X}_{a},\varphi,\boldsymbol{A}\right),\epsilon\in\mathbb{R}.\label{eq:PF-time-trans}
\end{equation}
The infinitesimal generator and characteristic are calculated as
\begin{align}
 & \boldsymbol{v}=\frac{\partial}{\partial t},\medspace\xi^{t}=1,\boldsymbol{\kappa}=0,\:\boldsymbol{\theta}_{a}=0,\medspace\phi_{0}=\boldsymbol{\phi}_{\boldsymbol{A}}=0,\label{eq:PF-inf-generator-time}\\
 & \boldsymbol{q}_{a}=-\dot{\boldsymbol{X}}_{a},\text{\ensuremath{\;}}Q_{0}=-\varphi_{,t},\medspace\boldsymbol{Q}_{\boldsymbol{A}}=-\boldsymbol{A}_{,t}.\label{eq:PF-cha-time}
\end{align}
And the infinitesimal criterion (\ref{eq:PF-infinitesimal-criterion})
of the symmetry is
\begin{equation}
\frac{\partial\mathcal{L}}{\partial t}=0.\label{eq:PF-inf-time-trans}
\end{equation}
The corresponding energy conservation law is thus
\begin{align}
 & \partial_{t}\left[\mathcal{L}+\frac{1}{c}\boldsymbol{A}_{,t}\cdot\boldsymbol{E}_{\boldsymbol{E}}\left(\mathcal{L}\right)-\sum_{a}\left(\dot{\boldsymbol{X}}_{a}\cdot\frac{\partial\mathcal{L}}{\partial\dot{\boldsymbol{X}}_{a}}\right)\right]+\boldsymbol{\nabla}\cdot\left\{ \left[\varphi_{,t}\boldsymbol{E}_{\boldsymbol{E}}\left(\mathcal{L}\right)-\boldsymbol{A}_{,t}\times\boldsymbol{E}_{\boldsymbol{B}}\left(\mathcal{L}\right)\right]\right.\nonumber \\
 & \left.\vphantom{\frac{\partial\mathcal{L}_{a}}{\partial\dot{\boldsymbol{X}}_{a}}}-\sum_{a}\left(\dot{\boldsymbol{X}}_{a}\frac{\partial\mathcal{L}_{a}}{\partial\dot{\boldsymbol{X}}_{a}}-\mathcal{L}_{a}\boldsymbol{I}\right)\cdot\dot{\boldsymbol{X}}_{a}\right\} +D_{\mu}\left[\mathcal{\mathbb{P}}_{1}^{\mu}+\mathcal{\mathbb{P}}_{2}^{\mu}\right]=0,\label{eq:PF-energy-conservation}
\end{align}
where

\begin{align}
 & \mathcal{\mathbb{P}}_{1}^{\mu}=\sum_{i=1}^{n}\sum_{j=1}^{i}\left(-1\right)^{j}D_{\mu_{j+1}}\cdots D_{\mu_{i}}\partial_{t}\boldsymbol{E}\cdot\left[D_{\mu_{1}}\cdots D_{\mu_{j-1}}\frac{\partial\mathcal{L}}{\partial D_{\mu_{1}}\cdots D_{\mu_{j-1}}D_{\mu}D_{\mu_{j+1}}\cdots D_{\mu_{i}}\boldsymbol{E}}\right],\label{eq:PF-energy-p1}\\
 & \mathcal{\mathbb{P}}_{2}^{\mu}=\sum_{i=1}^{n}\sum_{j=1}^{i}\left(-1\right)^{j}D_{\mu_{j+1}}\cdots D_{\mu_{i}}\partial_{t}\boldsymbol{B}\cdot\left[D_{\mu_{1}}\cdots D_{\mu_{j-1}}\frac{\partial\mathcal{L}}{\partial D_{\mu_{1}}\cdots D_{\mu_{j-1}}D_{\mu}D_{\mu_{j+1}}\cdots D_{\mu_{i}}\boldsymbol{B}}\right].\label{eq:PF-energy-p2}
\end{align}
The energy density and flux in Eq.\,(\ref{eq:PF-energy-conservation})
are obviously gauge dependent. To gauge-symmetrize the conservation
law, we add Eq.\,(\ref{eq:PF-identity-1}) to Eq.\,(\ref{eq:PF-energy-conservation})
and obtain,
\begin{align}
 & \partial_{t}\left[\mathcal{L}-\frac{\partial\mathcal{L}}{\partial\varphi}\varphi-\sum_{a}\left(\dot{\boldsymbol{X}}_{a}\cdot\frac{\partial\mathcal{L}}{\partial\dot{\boldsymbol{X}}_{a}}\right)-\boldsymbol{E}\cdot\boldsymbol{E}_{\boldsymbol{E}}\left(\mathcal{L}\right)\right]+\boldsymbol{\nabla}\cdot\left\{ c\boldsymbol{E}\times\boldsymbol{E}_{\boldsymbol{B}}\left(\mathcal{L}\right)\right.\nonumber \\
 & \left.\vphantom{\frac{\partial\mathcal{L}_{a}}{\partial\dot{\boldsymbol{X}}_{a}}}+c\frac{\partial\mathcal{L}}{\partial\boldsymbol{A}}\varphi-\sum_{a}\left(\dot{\boldsymbol{X}}_{a}\frac{\partial\mathcal{L}_{a}}{\partial\dot{\boldsymbol{X}}_{a}}-\mathcal{L}_{a}\boldsymbol{I}\right)\cdot\dot{\boldsymbol{X}}_{a}\right\} +D_{\mu}\left[\mathcal{\mathbb{P}}_{1}^{\mu}+\mathcal{\mathbb{P}}_{2}^{\mu}\right]=0.\label{eq:PF-energy-conser-gauge-inv}
\end{align}

In deriving Eq.\,(\ref{eq:PF-energy-conser-gauge-inv}), we have
rewritten the first and second terms of Eq.\,(\ref{eq:PF-identity-1})
as 
\begin{align}
 & \frac{D}{Dt}\left\{ \frac{D}{D\boldsymbol{x}}\cdot\left[\boldsymbol{E}_{\boldsymbol{E}}\left(\mathcal{L}\right)\varphi\right]\right\} =\frac{D}{Dt}\left\{ \boldsymbol{\nabla}\cdot\left[\boldsymbol{E}_{\boldsymbol{E}}\left(\mathcal{L}\right)\right]\varphi+\boldsymbol{E}_{\boldsymbol{E}}\left(\mathcal{L}\right)\cdot\boldsymbol{\nabla}\varphi\right\} \nonumber \\
 & =\frac{D}{Dt}\left\{ -\frac{\partial\mathcal{L}}{\partial\varphi}\varphi+\boldsymbol{E}_{\boldsymbol{E}}\left(\mathcal{L}\right)\cdot\boldsymbol{\nabla}\varphi\right\} ,\label{eq:PF-identity-1-1-trans}
\end{align}
\begin{align}
 & \frac{D}{D\boldsymbol{x}}\cdot\left\{ \frac{D}{Dt}\left[-\boldsymbol{E}_{\boldsymbol{E}}\left(\mathcal{L}\right)\varphi\right]\right\} =\frac{D}{D\boldsymbol{x}}\cdot\left\{ -\frac{\partial}{\partial t}\left[\boldsymbol{E}_{\boldsymbol{E}}\left(\mathcal{L}\right)\right]\varphi-\boldsymbol{E}_{\boldsymbol{E}}\left(\mathcal{L}\right)\varphi_{,t}\right\} \nonumber \\
 & =\frac{D}{D\boldsymbol{x}}\cdot\left\{ c\boldsymbol{\nabla}\times\left[\boldsymbol{E}_{\boldsymbol{B}}\left(\mathcal{L}\right)\right]\varphi+c\frac{\partial\mathcal{L}}{\partial\boldsymbol{A}}\varphi-\boldsymbol{E}_{\boldsymbol{E}}\left(\mathcal{L}\right)\varphi_{,t}\right\} \nonumber \\
 & =\frac{D}{D\boldsymbol{x}}\cdot\left\{ -\boldsymbol{E}_{\boldsymbol{E}}\left(\mathcal{L}\right)\varphi_{,t}-c\boldsymbol{\nabla}\varphi\times\boldsymbol{E}_{\boldsymbol{B}}\left(\mathcal{L}\right)+c\frac{\partial\mathcal{L}}{\partial\boldsymbol{A}}\varphi\right\} ,\label{eq:PF-identity-1-2-trans}
\end{align}
where use has been made of Eqs.\,(\ref{eq:PF-Maxwell's-Eq1}) and
(\ref{eq:PF-Maxwell's-Eq2}). Adding Eqs.\,(\ref{eq:PF-identity-1-1-trans})
and (\ref{eq:PF-identity-1-2-trans}) to Eq.\,(\ref{eq:PF-energy-conservation})
leads to Eq.\,(\ref{eq:PF-energy-conser-gauge-inv}). 

We now prove that the energy density and flux in Eq.\,(\ref{eq:PF-energy-conser-gauge-inv})
are gauge symmetric. It suffices to show that the following terms
\begin{equation}
\begin{alignedat}{1}s_{1} & \equiv\mathcal{L}-\frac{\partial\mathcal{L}}{\partial\varphi}\varphi-\sum_{a}\left(\dot{\boldsymbol{X}}_{a}\cdot\frac{\partial\mathcal{L}}{\partial\dot{\boldsymbol{X}}_{a}}\right),\\
\boldsymbol{s}_{2} & \equiv c\frac{\partial\mathcal{L}}{\partial\boldsymbol{A}}\varphi-\sum_{a}\left(\dot{\boldsymbol{X}}_{a}\frac{\partial\mathcal{L}_{a}}{\partial\dot{\boldsymbol{X}}_{a}}-\mathcal{L}_{a}\boldsymbol{I}\right)\cdot\dot{\boldsymbol{X}}_{a}
\end{alignedat}
\label{eq:S1+S2}
\end{equation}
are gauge symmetric. Substituting Eq.\,(\ref{eq:PF-Lag-den-2-gauge-seperated})
into the expression of $s_{1}$, we have 
\begin{align}
 & s_{1}=\sum_{a}\left[-q_{a}\varphi\delta_{a}+\frac{q_{a}}{c}\boldsymbol{A}\cdot\dot{\boldsymbol{X}}_{a}\delta_{a}\right]+\text{GIP}\left(\mathcal{L}\right)\nonumber \\
 & -\frac{\partial}{\partial\varphi}\left[\sum_{a}\left(-q_{a}\varphi\delta_{a}+\frac{q_{a}}{c}\boldsymbol{A}\cdot\dot{\boldsymbol{X}}_{a}\delta_{a}\right)\right]\varphi-\sum_{a}\left[\dot{\boldsymbol{X}}_{a}\cdot\frac{\partial}{\partial\dot{\boldsymbol{X}}_{a}}\left(-q_{a}\varphi\delta_{a}+\frac{q_{a}}{c}\boldsymbol{A}\cdot\dot{\boldsymbol{X}}_{a}\delta_{a}\right)\right]\nonumber \\
 & =\text{GSP}\left(\mathcal{L}\right).\label{eq:s1-proof}
\end{align}
Similarly, $\boldsymbol{s}_{2}$ is also gauge symmetric, 
\begin{align}
 & \boldsymbol{s}_{2}=c\frac{\partial\mathcal{L}}{\partial\boldsymbol{A}}\varphi-\sum_{a}\left(\dot{\boldsymbol{X}}_{a}\frac{\partial\mathcal{L}_{a}}{\partial\dot{\boldsymbol{X}}_{a}}-\mathcal{L}_{a}\boldsymbol{I}\right)\cdot\dot{\boldsymbol{X}}_{a}=c\frac{\partial}{\partial\boldsymbol{A}}\left[\sum_{a}\left(-q_{a}\varphi\delta_{a}+\frac{q_{a}}{c}\boldsymbol{A}\cdot\dot{\boldsymbol{X}}_{a}\delta_{a}\right)\right]\varphi\nonumber \\
 & -\sum_{a}\left\{ \dot{\boldsymbol{X}}_{a}\frac{\partial}{\partial\dot{\boldsymbol{X}}_{a}}\left(-q_{a}\varphi\delta_{a}+\frac{q_{a}}{c}\boldsymbol{A}\cdot\dot{\boldsymbol{X}}_{a}\delta_{a}\right)-\left(-q_{a}\varphi\delta_{a}+\frac{q_{a}}{c}\boldsymbol{A}\cdot\dot{\boldsymbol{X}}_{a}\delta_{a}\right)\boldsymbol{I}+\text{GSP}\left(\mathcal{L}_{a}\right)\boldsymbol{I}\right\} \cdot\dot{\boldsymbol{X}}_{a}\nonumber \\
 & =\sum_{a}\text{GSP}\left(\mathcal{L}_{a}\right)\dot{\boldsymbol{X}}_{a}.\label{eq:s2-proof}
\end{align}

\subsection{Gauge-symmetric momentum conservation law \label{subsec:PF-The-improvement-momentum}}

We now discuss how to derive a gauge-symmetric momentum conservation
law, assuming that the action $\mathcal{A}\equiv\int\mathcal{L}dtd^{3}\boldsymbol{x}$
of the electromagnetic field-charged particle system is invariant
under the space translation
\begin{equation}
\left(t,x,\boldsymbol{X}_{a},\varphi,\boldsymbol{A}\right)\mapsto\left(t,\boldsymbol{x}+\epsilon\boldsymbol{h},\boldsymbol{X}_{a}+\epsilon\boldsymbol{h},\varphi,\boldsymbol{A}\right),\epsilon\in\mathbb{R}.\label{eq:PF-space-translation}
\end{equation}
We emphasize that, different from the situation in standard field
theories, this symmetry group simultaneously translates both the spatial
coordinate $\boldsymbol{x}$ for the field and particle's position
$\boldsymbol{X}_{a}$ \citep{Qin2014b,Fan2018,Fan2019}. The infinitesimal
criterion of this symmetry is 
\begin{equation}
\frac{\partial\mathcal{L}}{\partial\boldsymbol{x}}+\sum_{a}\frac{\partial\mathcal{L}}{\partial\boldsymbol{X}_{a}}=0.\label{eq:PF-space-trans-inf-criterion}
\end{equation}
From Eq.\,(\ref{eq:PF-space-translation}), the infinitesimal generator
and its characteristic are
\begin{align}
 & \boldsymbol{v}=\boldsymbol{h}\cdot\sum_{a}\left(\frac{\partial}{\partial\boldsymbol{x}}+\frac{\partial}{\partial\boldsymbol{X}_{a}}\right),\medspace\xi^{t}=0,\boldsymbol{\kappa}=\boldsymbol{\theta}_{a}=\boldsymbol{h},,\medspace\phi_{0}=\boldsymbol{\phi}_{\boldsymbol{A}}=0,\label{eq:PF-space-trans-generator}\\
 & \boldsymbol{q}_{a}=\boldsymbol{h},Q_{0}=-\boldsymbol{h}\cdot\boldsymbol{\nabla}\varphi,\medspace\boldsymbol{Q}_{\boldsymbol{A}}=-\boldsymbol{h}\cdot\boldsymbol{\nabla}\boldsymbol{A}.\label{eq:PF-space-trans-charact}
\end{align}
The corresponding momentum conservation law is obtained by substituting
Eqs.\,(\ref{eq:PF-space-trans-generator}) and (\ref{eq:PF-space-trans-charact})
into Eq.\,(\ref{eq:PF-general-conservation}), i.e., 
\begin{align}
 & \partial_{t}\left[\frac{1}{c}\boldsymbol{E}_{\boldsymbol{E}}\left(\mathcal{L}\right)\cdot\left(\boldsymbol{\nabla}\boldsymbol{A}\right)^{T}+\sum_{a}\left(\frac{\partial\mathcal{L}}{\partial\dot{\boldsymbol{X}}_{a}}\right)\right]+\boldsymbol{\nabla}\cdot\left\{ \mathcal{L}\boldsymbol{I}+\sum_{a}\left(\dot{\boldsymbol{X}}_{a}\frac{\partial\mathcal{L}_{a}}{\partial\dot{\boldsymbol{X}}_{a}}-\mathcal{L}_{a}\boldsymbol{I}\right)\right.\nonumber \\
 & \left.\vphantom{\frac{\partial\mathcal{L}_{a}}{\partial\dot{\boldsymbol{X}}_{a}}}+\left[\boldsymbol{E}_{\boldsymbol{E}}\left(\mathcal{L}\right)\boldsymbol{\nabla}\varphi+\boldsymbol{E}_{\boldsymbol{B}}\left(\mathcal{L}\right)\times\left(\boldsymbol{\nabla}\boldsymbol{A}\right)^{T}\right]\right\} +D_{\mu}\left[\mathcal{\bar{\mathbb{\boldsymbol{P}}}}_{1}^{\mu}+\mathcal{\bar{\mathbb{P}}}_{2}^{\mu}\right]=0,\label{eq:PF-momentum-conservation}
\end{align}
where
\begin{align}
 & \mathcal{\bar{\mathbb{\boldsymbol{P}}}}_{1}^{\mu}=\sum_{i=1}^{n}\sum_{j=1}^{i}\left(-1\right)^{j}D_{\mu_{j+1}}\cdots D_{\mu_{i}}\boldsymbol{\nabla}\boldsymbol{E}\cdot\left[D_{\mu_{1}}\cdots D_{\mu_{j-1}}\frac{\partial\mathcal{L}}{\partial D_{\mu_{1}}\cdots D_{\mu_{j-1}}D_{\mu}D_{\mu_{j+1}}\cdots D_{\mu_{i}}\boldsymbol{E}}\right],\label{eq:PF-momentum-p1}\\
 & \mathcal{\bar{\mathbb{\boldsymbol{P}}}}_{2}^{\mu}=\sum_{i=1}^{n}\sum_{j=1}^{i}\left(-1\right)^{j}D_{\mu_{j+1}}\cdots D_{\mu_{i}}\boldsymbol{\nabla}\boldsymbol{B}\cdot\left[D_{\mu_{1}}\cdots D_{\mu_{j-1}}\frac{\partial\mathcal{L}}{\partial D_{\mu_{1}}\cdots D_{\mu_{j-1}}D_{\mu}D_{\mu_{j+1}}\cdots D_{\mu_{i}}\boldsymbol{B}}\right].\label{eq:PF-momentum-p2}
\end{align}
Again, the momentum density and flux in Eq.\,(\ref{eq:PF-momentum-conservation})
are gauge dependent. We add Eq.\,(\ref{eq:PF-identity-2}) to Eq.\,(\ref{eq:PF-momentum-conservation})
to obtain a gauge-symmetric momentum conservation law, 
\begin{align}
 & \partial_{t}\left[\frac{1}{c}\boldsymbol{E}_{\boldsymbol{E}}\left(\mathcal{L}\right)\times\boldsymbol{B}+\frac{1}{c}\frac{\partial\mathcal{L}}{\partial\varphi}\boldsymbol{A}+\sum_{a}\left(\frac{\partial\mathcal{L}}{\partial\dot{\boldsymbol{X}}_{a}}\right)\right]+\boldsymbol{\nabla}\cdot\left\{ -\frac{\partial\mathcal{L}}{\partial\boldsymbol{A}}\boldsymbol{A}+\sum_{a}\left(\dot{\boldsymbol{X}}_{a}\frac{\partial\mathcal{L}_{a}}{\partial\dot{\boldsymbol{X}}_{a}}\right)\right.\nonumber \\
 & \left.\vphantom{\frac{\partial\mathcal{L}}{\partial\boldsymbol{A}_{\alpha}}}\left[\mathcal{L}_{F}-\boldsymbol{B}\cdot\boldsymbol{E}_{\boldsymbol{B}}\left(\mathcal{L}\right)\right]\boldsymbol{I}+\left[-\boldsymbol{E}_{\boldsymbol{E}}\left(\mathcal{L}\right)\boldsymbol{E}+\boldsymbol{B}\boldsymbol{E}_{\boldsymbol{B}}\left(\mathcal{L}\right)\right]\right\} +D_{\mu}\left[\left(\mathcal{\bar{\mathbb{\boldsymbol{P}}}}_{1}^{\mu}+\mathcal{\bar{\mathbb{\boldsymbol{P}}}}_{2}^{\mu}\right)\right]=0.\label{eq:PF-momentum-conser-gauge-inv}
\end{align}
In the derivation of Eq.\,(\ref{eq:PF-momentum-conser-gauge-inv}),
we have rewritten the first and second terms of Eq.\,(\ref{eq:PF-identity-2})
as
\begin{align}
 & \frac{D}{Dt}\left\{ \frac{D}{D\boldsymbol{x}}\cdot\left[-\frac{1}{c}\boldsymbol{E}_{\boldsymbol{E}}\left(\mathcal{L}\right)\boldsymbol{A}\right]\right\} =\frac{D}{Dt}\left\{ -\frac{1}{c}\boldsymbol{\nabla}\cdot\left[\boldsymbol{E}_{\boldsymbol{E}}\left(\mathcal{L}\right)\right]\boldsymbol{A}-\frac{1}{c}\boldsymbol{E}_{\boldsymbol{E}}\left(\mathcal{L}\right)\cdot\boldsymbol{\nabla}\boldsymbol{A}\right\} \nonumber \\
 & =\frac{D}{Dt}\left[\frac{1}{c}\frac{\partial\mathcal{L}}{\partial\varphi}\boldsymbol{A}-\frac{1}{c}\boldsymbol{E}_{\boldsymbol{E}}\left(\mathcal{L}\right)\cdot\boldsymbol{\nabla}\boldsymbol{A}\right],\label{eq:PF-identity-2-1}
\end{align}
\begin{align}
 & \frac{D}{D\boldsymbol{x}}\cdot\left\{ \frac{D}{Dt}\left[\frac{1}{c}\boldsymbol{E}_{\boldsymbol{E}}\left(\mathcal{L}\right)\boldsymbol{A}\right]\right\} =\frac{D}{D\boldsymbol{x}}\cdot\left\{ \frac{1}{c}\frac{\partial}{\partial t}\left[\boldsymbol{E}_{\boldsymbol{E}}\left(\mathcal{L}\right)\right]\boldsymbol{A}+\frac{1}{c}\boldsymbol{E}_{\boldsymbol{E}}\left(\mathcal{L}\right)\boldsymbol{A}_{,t}\right\} \nonumber \\
 & =\frac{D}{D\boldsymbol{x}}\cdot\left\{ -\boldsymbol{\nabla}\times\left[\boldsymbol{E}_{\boldsymbol{B}}\left(\mathcal{L}\right)\right]\boldsymbol{A}-\frac{\partial\mathcal{L}}{\partial\boldsymbol{A}}\boldsymbol{A}+\frac{1}{c}\boldsymbol{E}_{\boldsymbol{E}}\left(\mathcal{L}\right)\boldsymbol{A}_{,t}\right\} \nonumber \\
 & =\frac{D}{D\boldsymbol{x}}\cdot\left\{ -\boldsymbol{E}_{\boldsymbol{B}}\left(\mathcal{L}\right)\times\boldsymbol{\nabla}\boldsymbol{A}-\frac{\partial\mathcal{L}}{\partial\boldsymbol{A}}\boldsymbol{A}+\frac{1}{c}\boldsymbol{E}_{\boldsymbol{E}}\left(\mathcal{L}\right)\boldsymbol{A}_{,t}\right\} ,\label{eq:PF-identity-2-2}
\end{align}
where use has been made of Eqs.\,(\ref{eq:PF-Maxwell's-Eq1}) and
(\ref{eq:PF-Maxwell's-Eq2}). Adding Eqs.\,(\ref{eq:PF-identity-2-1})
and (\ref{eq:PF-identity-2-2}) into Eq.\,(\ref{eq:PF-momentum-conservation})
gives Eq.\,(\ref{eq:PF-momentum-conser-gauge-inv}).

To show that the momentum density and flux in Eq.\,(\ref{eq:PF-momentum-conser-gauge-inv})
are gauge symmetric, it suffices to show that the following terms
are gauge symmetric,
\begin{equation}
\begin{alignedat}{1}\boldsymbol{t}_{1} & \equiv\frac{1}{c}\frac{\partial\mathcal{L}}{\partial\varphi}\boldsymbol{A}+\sum_{a}\left(\frac{\partial\mathcal{L}}{\partial\dot{\boldsymbol{X}}_{a}}\right),\\
\boldsymbol{t}_{2} & \equiv-\frac{\partial\mathcal{L}}{\partial\boldsymbol{A}}\boldsymbol{A}+\sum_{a}\left(\dot{\boldsymbol{X}}_{a}\frac{\partial\mathcal{L}_{a}}{\partial\dot{\boldsymbol{X}}_{a}}\right).
\end{alignedat}
\label{eq:t1+t2}
\end{equation}
Substituting Eq.\,(\ref{eq:PF-Lag-den-2-gauge-seperated}) into the
expression of $\boldsymbol{t}_{1},$ we can see that it is gauge symmetric,
i.e., 
\begin{align}
 & \boldsymbol{t}_{1}=\frac{1}{c}\frac{\partial\mathcal{L}}{\partial\varphi}\boldsymbol{A}+\sum_{a}\left(\frac{\partial\mathcal{L}}{\partial\dot{\boldsymbol{X}}_{a}}\right)=\frac{1}{c}\sum_{a}\frac{\partial}{\partial\varphi}\left(-q_{a}\varphi\delta_{a}+\frac{q_{a}}{c}\boldsymbol{A}\cdot\dot{\boldsymbol{X}}_{a}\delta_{a}\right)\boldsymbol{A}+\sum_{a}\frac{\partial}{\partial\dot{\boldsymbol{X}}_{a}}\left[-q_{a}\varphi\delta_{a}\right.\nonumber \\
 & \left.\vphantom{\left(\mathcal{L}_{a}\right)}+\frac{q_{a}}{c}\boldsymbol{A}\cdot\dot{\boldsymbol{X}}_{a}\delta_{a}+\text{GSP}\left(\mathcal{L}_{a}\right)\right]=\sum_{a}\frac{\partial}{\partial\dot{\boldsymbol{X}}_{a}}\left[\text{GSP}\left(\mathcal{L}_{a}\right)\right].\label{eq:t1}
\end{align}
Similarly, $\boldsymbol{t}_{2}$ is also gauge-symmetric, 
\begin{align}
 & \boldsymbol{t}_{2}=-\frac{\partial\mathcal{L}}{\partial\boldsymbol{A}}\boldsymbol{A}+\sum_{a}\left(\dot{\boldsymbol{X}}_{a}\frac{\partial\mathcal{L}_{a}}{\partial\dot{\boldsymbol{X}}_{a}}\right)=-\frac{\partial}{\partial\boldsymbol{A}}\left(-q_{a}\varphi\delta_{a}+\frac{q_{a}}{c}\boldsymbol{A}\cdot\dot{\boldsymbol{X}}_{a}\delta_{a}\right)\boldsymbol{A}\nonumber \\
 & +\sum_{a}\dot{\boldsymbol{X}}_{a}\frac{\partial}{\partial\dot{\boldsymbol{X}}_{a}}\left[-q_{a}\varphi\delta_{a}+\frac{q_{a}}{c}\boldsymbol{A}\cdot\dot{\boldsymbol{X}}_{a}\delta_{a}+\text{GSP}\left(\mathcal{L}_{a}\right)\right]=\sum_{a}\dot{\boldsymbol{X}}_{a}\frac{\partial}{\partial\dot{\boldsymbol{X}}_{a}}\left[\text{GSP}\left(\mathcal{L}_{a}\right)\right].\label{eq:t2}
\end{align}

\section{conclusion}

In this study, we developed a gauge-symmetrization method for the
energy and momentum conservation laws in general high-order classical
electromagnetic field theories, which appear in the study of gyrokinetic
systems \citep{Qin2005,Qin2007a,Fan2020} for magnetized plasmas and
the Podolsky system \citep{Bopp1940,Podolsky1942} for the radiation
reaction of classical charged particles. The method only removes the
electromagnetic gauge dependence from the canonical EMT derived from
the spacetime translation symmetry, without necessarily symmetrizing
the EMT with respect to the tensor indices. This is adequate for applications
not involving general relativity.

To achieve this goal, we reformulated the EL equation and infinitesimal
criterion in terms of the Faraday tensor $F_{\mu\nu}$. The canonical
EMT $T_{\text{N}}^{\mu\nu}$ is derived using this formalism, and
it was found that the gauge dependent part of $T_{\text{N}}^{\mu\nu}$
can be removed by adding the divergence of the displacement-potential
tensor, which is defined as 
\begin{equation}
\begin{aligned} & \mathcal{F}^{\sigma\mu\nu}\equiv\frac{1}{4\pi}\mathcal{D}^{\sigma\mu}A^{\nu}.\end{aligned}
\label{eq:155}
\end{equation}
It was shown that the displacement-potential tensor $\mathcal{F}^{\sigma\mu\nu}$
is related to the well-known BR super-potential $\mathcal{S}^{\sigma\mu\nu}$
as
\begin{equation}
\mathcal{S}^{\sigma\mu\nu}=\mathcal{F}^{\sigma\mu\nu}+\frac{1}{2}\left[\Delta^{\sigma\nu\mu}-\Delta^{\mu\nu\sigma}-\Delta^{\nu\mu\sigma}\right],\label{eq:157}
\end{equation}
where $\Delta^{\sigma\nu\mu}$ is defined in Eq.\,(\ref{eq:cov-Delta}).
Using the example of the Podolsky system \citep{Bopp1940,Podolsky1942},
we show that $\Delta^{\sigma\nu\mu}$ in general is non-vanishing
for high-order field theories. For a first-order field theory, such
as the standard Maxwell system (\ref{eq:Maxwell-Lagrangian}), $\Delta^{\sigma\nu\mu}$
vanishes such that $\mathcal{S}^{\sigma\mu\nu}=\mathcal{F}^{\sigma\mu\nu}$.
In the case, the method developed can be used as a simpler procedure
to calculate the BR super-potential $\mathcal{S}^{\sigma\mu\nu}$
without the necessity to calculate the angular momentum tensor in
4D spacetime. 

Lastly, we applied the method to derive gauge-symmetric EMTs for high-order
electromagnetic systems coupled with classical charged particles.
Using the ``3+1'' form of Eq.\,(\ref{eq:cov-identity}), we obtained
the explicitly gauge-symmetric energy and momentum conservation laws
in a general setting {[}see Eqs.\,(\ref{eq:PF-energy-conser-gauge-inv})
and (\ref{eq:PF-momentum-conser-gauge-inv}){]}.
\begin{acknowledgments}
P. Fan was supported by Shenzhen Clean Energy Research Institute and
National Natural Science Foundation of China (NSFC-12005141). J. Xiao
was supported by the National MC Energy R\&D Program (2018YFE0304100),
National Key Research and Development Program (2016YFA0400600, 2016YFA0400601
and 2016YFA0400602), and the National Natural Science Foundation of
China (NSFC-11905220 and 11805273). H. Qin was supported by the U.S.
Department of Energy (DE-AC02-09CH11466).
\end{acknowledgments}

\bibliographystyle{apsrev4-1}
\bibliography{GaugeEliminationProcess}

\end{document}